\documentclass[twocolumn,showpacs,preprintnumbers,amsmath,amssymb]{revtex4} 
\usepackage{graphics}

\newcommand\beq{\begin{equation}}
\newcommand\eeq{\end{equation}}
\newcommand\bea{\begin{eqnarray}}
\newcommand\eea{\end{eqnarray}}

\begin{document}
\draft
\title { Fermionic ground state at unitarity and Haldane Exclusion 
Statistics}
\author{
R.K. Bhaduri$^{1,2}$, M.V.N. Murthy$^{1}$ and M. Brack$^3$ 
} 
\affiliation{1. The Institute of Mathematical Sciences, Chennai 600113, 
India}
\affiliation{2. Department of Physics and Astronomy, McMaster University,
Hamilton, Canada L8S 4M1}

\affiliation{3. Institute for Theoretical Physics, University of 
Regensburg, Regensburg, Germany} 
\date{\today}

\begin{abstract}

We consider a few-particle system of trapped neutral fermionic atoms at 
ultra-low temperatures, with the attractive interaction tuned to 
Feshbach resonance. We calculate the energies and the spatial densities 
of the few-body systems using a generalisation of the extended 
Thomas-Fermi (ETF) method, and assuming the particles obey the 
Haldane-Wu fractional exclusion statistics (FES) at unitarity. This 
method is different from the scaled ETF version given by Chang and 
Bertsch (Phys. Rev. A{\bf 76},021603(R) (2007)). Our semiclassical FES 
results are consistent with the Monte-Carlo calculations of the above 
authors, but can hardly be distinguished from their over all scaling of 
the ETF result at unitarity.

\vspace{0.5cm}
\end{abstract}
\maketitle

\section{Introduction} 

There has been a lot of interest in a dilute gas of neutral fermionic 
atoms at ultra-cold temperatures both experimentally~\cite{regal} and 
theoretically~\cite{baker}. In general, the low-energy properties of the 
gas are determined by the scattering length $a$, the number density $n$, 
and the temperature $T$ of the gas ( the effective range $r_0$ is small, 
so that $r_0/|a| \rightarrow 0$ as $a$ becomes large ). When the 
attractive interaction between the atoms is increased continuously by 
magnetic tuning from weak to strong, the scattering length $a$ goes from 
a small negative to a small positive value. In between, there is a 
zero-energy two-body bound state, and $|a|$ is infinite. The gas is said 
to be at unitarity in this situation, and the length scale $a$ drops 
out. The behaviour of the gas is expected to be universal at 
unitarity~\cite{baker}.  Understanding the ground state of the system in 
this limit is a challenge to many-body theorists as originally discussed 
by Bertsch~\cite{bertsch0}. Experimentally, if the temperature is small 
enough, a BCS superfluid is observed at the weak end, and a BEC 
condensate of dimers at the strong end~\cite{regal}. This was predicted 
long back by Leggett~\cite{leggett}, who extended the BCS formalism in a 
novel fashion to analyse the physical situation.  The BCS to BEC 
transition is found to be smooth, with no discontinuity in properties 
across the unitary point.

Chang and Bertsch~\cite{bertsch} have recently presented an {\it ab 
initio} Green's Function Monte-Carlo (GFMC) calculation of the energy 
and density of $N=2-22$ trapped fermionic atoms in a harmonic potential. 
The atoms are interacting via a short range central two-body potential, 
with its strength adjusted to yield a zero-energy two-body bound state 
in free space. The many-body properties of this system are expected to 
be independent of the shape of the two-body interaction in such a 
set-up.

In an earlier paper~\cite{bhaduri} we obtained the energy per particle 
and the chemical potential of a noninteracting gas of atoms at finite 
temperatures obeying fractional exclusion statistics 
(FES)~\cite{haldane,wu}. We assumed that at unitarity, the effect of the 
interaction could be simulated by FES for the bulk properties of the 
system. Our results, with the choice of one free parameter in FES, were 
found to be in good agreement with theoretical MC calculations for a 
free gas~\cite{bulag,burovski} and the experimental results in a trap 
\cite{kinast,hu}. Since the number of atoms, $N$, was taken to be large, 
no finite-N corrections were needed in our semiclassical calculations. 
This is not the case in the present paper, where $N$ is taken to be 
small. The purpose of this paper is to test whether the FES hypothesis 
gives improved results when the finite-N corrections are incorporated in 
our calculations.

We briefly recall the rationale for using FES. As is well known, FES is 
realised by the Calogero-Sutherland model~\cite{csm} in one dimension 
\cite{murthy}. In two dimensions, the kinetic and potential energy 
densities of fermions interacting with a zero-range potential scale as 
the square of the spatial density, and obey FES in the 
mean-field approximation~\cite{rajat}. A hint 
that Haldane statistics is also realised for cold atoms in three 
dimensions at unitarity comes from the observation that the total energy 
per atom of the gas may be obtained by scaling the kinetic energy term 
by a constant factor ~\cite{baker}. This is however a necessary, but not 
a sufficient condition. A further hint comes from the fact that the 
second virial coefficient of the gas at unitarity is temperature 
independent~\cite{mueller}.  In exclusion statistics, the 
scale-invariant interaction between atoms alters the ideal Fermi (Bose) 
values of the (exchange) second virial coefficient $+(-) 2^{-5/2}$ by 
adding an interacting part~\cite{shankar}. When FES is incorporated in 
the $T=0$ Thomas-Fermi (TF) method, it gives the same expression as the 
scaled density functional approach of Papenbrock\cite{pap}. This 
constitutes the bulk of the smooth part of the energy. It is the small 
next order term in the Extended Thomas-Fermi (ETF) expression that 
behaves differently with scaling or FES. The former was done by Chang 
and Bertsch~\cite{bertsch}. In this paper, we incorporate FES in an 
improved version of ETF to test if it can differentiate between the two 
alternatives when compared with the GFMC results.

In the next section, we first summarise the TF and ETF results. Both 
these have limitations at the classical turning point, where the spatial 
density behaves discontinuously. To rectify this, we make use of a 
modified semiclassical method~\cite{old} that gives a continuous 
variation of the density across the turning point. Our semiclassical 
results incorporating FES are next compared with the GFMC calculations 
of Chang and Bertsch~\cite{bertsch} for fermions trapped in a three 
dimensional oscillator potential. We find that FES results are 
consistent with the many-body GFMC results.

\section{semiclassical Calculations incorporating FES}     

There has been much interest amongst theorists to calculate the 
properties of a gas in the unitary regime ($k_f|a|>>1$), where 
$k_f=(3\pi^2n)^{1/3}$ is the Fermi wave number of the noninteracting 
gas. This is a challenging task, since there is no small expansion 
parameter, and a perturbative calculation cannot be done.  In 
particular, at $T=0$, the energy per particle of the gas is calculated 
to be $ \frac{E}{N}=\xi~\frac{3}{5} \frac{\hbar^2 
k_f^2}{2M}$, where $\xi$ is a constant scaling factor~\cite{baker}. A MC 
calculation 
gives $\xi\simeq0.44$~\cite{carlson}. The experimental value is about 
0.5, but with large error bars~\cite{bart}. A similar relation may be 
obtained assuming a noninteracting gas obeying FES, with a statistical 
occupancy factor $g$. At $T=0$, FES gives 
$$N=V \frac{1}{g} \frac{2}{(2\pi)^3}\int_0^{\tilde{k_f}} 4\pi k^2 dk~,$$ 
where we have 
included a spin degeneracy factor of 2 and the factor $1/g$ is the FES 
occupancy factor at $T=0$ with $g=1$ as the fermionic limit.
The modified Fermi momentum 
$\tilde{k_f}$, from above. is $\tilde{k_f}=g^{1/3} k_f$, where $k_f$ is 
the fermi momentum of the noninteracting Fermi gas. It also follows that 
the energy per particle of the unitary gas is given by 
$\frac{E}{N}=g^{2/3}~\frac{3}{5}\frac{\hbar^2 k_f^2}{2M}~$. Comparing 
with the scaled version, we see that the scaling factor $\xi$ in a Fermi 
gas is related to the statistical parameter $g$ by the relation 
$\xi=g^{2/3}$. In a three-dimensional isotropic harmonic trap, a similar 
scaling of the TF expression gives~\cite{pap}, in units of 
$\hbar\omega$, 
\beq E_{TF}=\frac{\xi^{1/2}}{4}(3N)^{4/3}~. 
\label{TF1} 
\eeq 
In FES, an identical relation is obtained, with $\xi^{1/2}$ 
replaced by $g^{1/3}$. The scaled TF spatial density is also identical 
to the FES expression when this replacement is made : 
\beq 
\rho_{TF}(r)=\frac{1}{3\pi^2 g}(\frac{2}{l^2})^{3/2}\left((3gN)^{1/3}- 
\frac{1}{2}\frac{r^2}{l^2}\right)^{3/2}~, 
\eeq 
where 
$l=\sqrt{(\hbar/m\omega)}$. The above expression is valid for $r\leq 
r_0$, 
where $r_0=\sqrt{2}l(3gN)^{1/6}$ is the classical turning point. For 
$r>r_0$, the TF density is zero.  To implement finite-N corrections, one 
has to consider ETF~\cite{jennings}. Chang and 
Bertsch~\cite{bertsch} scale the energy expression for ETF by the same 
over all factor as in TF (denoted by ETF'), where as FES yields a 
different 
expression~\cite{brandon} (in units of $\hbar\omega$): 
\bea 
E_{ETF}'&=&\xi^{1/2}\left(\frac{(3N)^{4/3}}{4}+\frac{(3N)^{2/3}}{8}+... 
\right);\label{once}\\ 
E_{ETF}&=&\left( 
g^{1/3}\frac{(3N)^{4/3}}{4}+g^{-1/3} 
\frac{(3N)^{2/3}}{8}+...\right)~
\label{twice}. 
\eea 
Although ETF gives a 
reasonable description of the smooth part of the energy, it fails to do 
so for the spatial density. In fact, the ETF density diverges at the 
turning point. To give a consistent description of both the energy and 
the spatial density, we adopt a method where a selective summation of 
the higher order gradient terms of the Wigner-Kirkwood series is 
made~\cite{old}. For a harmonic trapping potential $V(r)$, retaining 
terms up to third order in $\beta$, the Bloch density $C(r,\beta)$ 
incorporating FES is given by 
\bea C(r,\beta)&=&\frac{1}{4\pi^2 
g}(\frac{2m}{\hbar^2\beta})^{3/2} \nonumber\\ 
&&\left(1-\frac{\hbar^2\beta^2}{12 m}\right) \exp\left[-\beta V+\beta^3 
\frac{\hbar^2}{24 m}(\nabla V)^2\right]~. 
\label{bloch} 
\eea 
The spatial density is obtained by taking the inverse Laplace transform of 
$C(r,\beta)/\beta$ with respect to the chemical potential $\mu$, which 
we denote by $\tilde{\rho}={\cal L}^{-1}_{\mu} [C(r,\beta)/\beta]$. It 
is the cubic term in the exponent that makes the density continuous 
across the classical turning point. Similarly, the energy is given by 
$\tilde{E}=\mu N-{\cal L}^{-1}_{\mu}C(r,\beta)/\beta^2$. The inverse 
Laplace transformations are carried out by the saddle-point method. The 
quality of the approximation is tested by applying the method to $N$ 
noninteracting spin-1/2 fermions $(g=1)$ in a harmonic potential. 
The 
result for the energy is plotted as a function of $N$ is plotted in 
Fig.1. 
To facilitate the comparison, the TF energy is subtracted out from the 
quantum as well as the semiclassical results for $E_{ETF}$ and 
$\tilde{E}$. 
Note that the shell effects in the energy as well as 
the density are not reproduced in the semiclassical calculations. 
\begin{figure}[h] 
\scalebox{0.7}{\includegraphics*{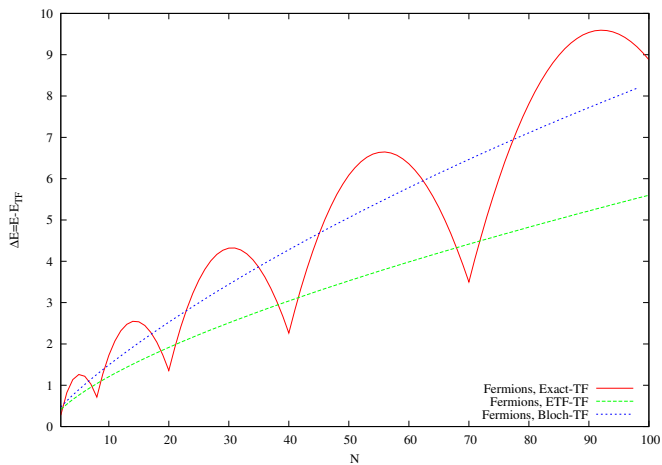}} 
\caption{Plot of the energy, $E-E_{TF}$  vs particle number for fermions 
(i.e., 
$g=1$). 
The red (solid) curve corresponds to the exact calculation in a harmonic 
oscillator while the 
green (dashed) and blue (dotted) curves correspond the calculations 
based on ETF and resummation methods.} 
\label{fig1} 
\end{figure}

In Fig.2, we compare the GFMC results~\cite{bertsch} of the energy for 
$N=2-22$ atoms with the various semiclassical calculations. For the latter,    
the scaling factor in Eq.(\ref{once}) is taken to be $\xi=0.48$, that 
corresponds to $g=1/3$ for the ETF Eq.(\ref{twice}), and also for $\tilde{E}$. 
Our choice of $\xi=0.48$ is very close to that of~\cite{bertsch}, and 
corresponds to a $g$ not too different from the value of $0.29$ chosen 
earlier~\cite{bhaduri}. It is seen from Fig.2 that all the semiclassical 
methods fare well, and it is not possible to distinguish the scaled results 
from the FES ones. 
\begin{figure}[h]
\scalebox{0.3}{\includegraphics*{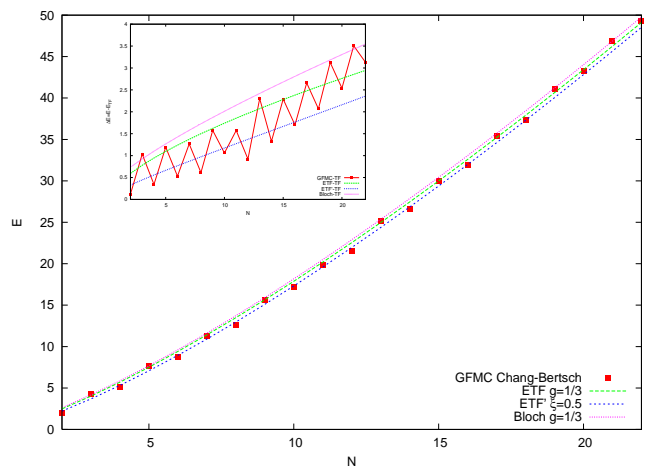}}
\caption{Plot of
the energy vs particle number with $g=1/3$. The data points
refer to the GFMC calculation of Chang and Bertsch\cite{bertsch}. The 
pink(dotted) curve corresponds to the energy calculation using the 
resummation 
method. The green (dashed) and blue(short-dashed) curves correspond to 
the standard ETF 
calculation with FES (see Eq.(\ref{twice})) and the ETF' approximation 
given by Chang and 
Bertsch (see eq.(\ref{once})). Inset shows the energy after subtracting 
the TF contribution 
as in Fig.1 but choosing $g=1/3$.} 
\label{fig2}
\end{figure}
A clearer comparison is made in the inset of Fig.2, where the large TF 
term given by Eq.(\ref{TF1}) is subtracted out from the energies. Even then, 
it is not possible to assert the relative superiority of over all scaling to 
FES. We suggest that a distinction may possibly be made if a larger range of 
$N$ values are spanned by a MC many-body calculation. An interesting 
aspect of GFMC results (see inset in Fig.2) is the odd-even oscillations 
in energy. 
%\begin{figure}[h]
%\scalebox{0.4}{\includegraphics*{geond.eps}}
%\caption{Plot of
%the energy vs particle number for fermions at $g=1/3$ after subtracting 
%the TF energy to show the differences more clearly. The data points 
%refer to the GFMC calculation of Chang and Bertsch\cite{bertsch}. The 
%cyon curve corresponds to the energy calculation using the resummation 
%method. The pink and blue curves correspond to the standard ETF 
%calculation with FES and the 'ETF' approximation given by Chang and 
%Bertsch. } 
%\label{fig3}
%\end{figure}
\begin{figure}[h]
\scalebox{0.7}{\includegraphics*{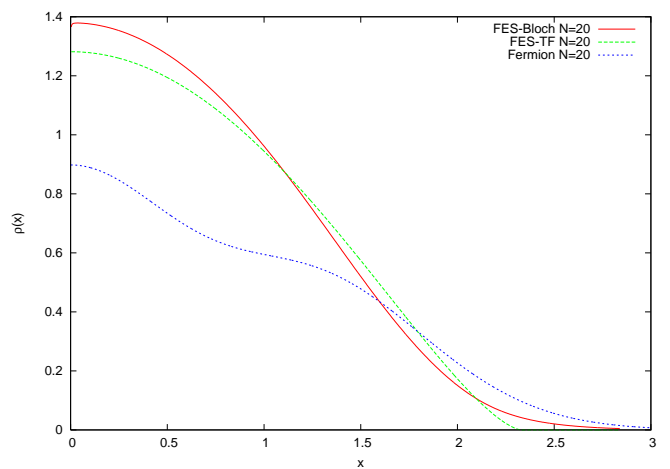}}
\caption{Plot of radial
density for $N=20$ as a function of the scaled distance $x$. Shown are 
the density calculated using resummation method (red-solid) and the TF 
density (green-dashed) with $g=1/3$. The density of fermions in a 
harmonic oscillator is also shown for comparison (blue-dotted).} 
\label{fig3}
\end{figure}
In Fig.3 the calculated density for $N=20$ particles is 
plotted using the resummation method and is compared with the TF density. 
Although there is not much to choose between the ETF and the resummation 
results for the energy, the density in the latter case is distinctly 
superior and  appears to agree 
with the smoothed part of the density calculated by the  
the  GFMC method. It also reproduces the tail beyond the turning point,
which is not 
possible in TF or ETF approximations.  

To summarise, We have considered a few particle system of trapped and 
interacting neutral fermionic atoms at ultra-low temperatures. The 
energy and spatial density of this system is calculated semiclassically 
assuming the particles obey the Haldane-Wu fractional exclusion 
statistics (FES) at unitarity.  The semiclassical FES results are 
consistent with the Monte-Carlo calculations of Chang and 
Bertsch~\cite{bertsch}, but can hardly be distinguished from the over 
all scaling of the noninteracting energy that is commonly used at 
unitarity. However, it is interesting to note that both at finite 
temperature~\cite{bhaduri} and at zero temperature the FES frame work 
yields reasonably good results.

R.K.B. and M.V.N.  acknowledge financial support of the Hans Vielberth 
University Foundation during their visits to Regensburg. M.B. 
acknowledges the hospitality of McMaster University and financial 
support by the NSERC.

\end{document}